\documentclass[a4paper,12pt]{article}
\usepackage{amssymb}
\usepackage{graphicx}
\usepackage{epsfig}
\begin{document}
\title{Transverse Momentum Dependent Parton Density Functions and a Self-Similarity based Model of Proton Structure Function  $ F_{2}(x,Q^{2}) $ at Large and Small $ x $}
\author{ $ \mathrm{Akbari \; Jahan}^{\star} $ and D K Choudhury\\ Department of Physics, Gauhati University,\\ Guwahati - 781 014, Assam, India.\\ $ {}^{\star} $Email: akbari.jahan@gmail.com}
\maketitle
\begin{abstract}
Unintegrated parton density functions (uPDFs) of Quantum Chromodynamics (QCD), also known as TMDPDFs, are generally used to study details of hadronic final states in high energy lepton-hadron and hadron-hadron collisions; while the integrated parton density functions (PDFs) are used for conventional deep inelastic inclusive processes. The self-similarity based Model of proton structure function $ F_{2}(x,Q^{2}) $ suggested in recent years are however based on specific uPDFs with self-similarity at small \textit{x}. In this work, we study large \textit{x} limit of such a Model and modify the defining uPDFs to make them compatible with theoretical expectations in such limit. Possibility of saturation of Froissart bound in this Model is discussed. We also reanalyze the PDFs in this approach using its conventional relation with TMDPDFs.\\\\\\\\
\end{abstract}
\section{Introduction}
Since nineteen eighties the notion of fractals has found its applicability in high energy physics through the self-similar nature of hadron multi-particle production processes \cite {1, 2, 3}. Specifically during mid-nineties, James D. Bjorken \cite {4} highlighted the fractality of parton cascades leading to the anomalous dimension of the phase space. Similarly in early twenties, fractal characters of hadrons had been pursued within a statistical quark model \cite {5} with considerable success.\\

Relevance of these ideas in the contemporary physics of deep inelastic scattering (DIS) was first noted by Dremin and Levtchenko \cite {6} in early nineties where it was shown that saturation of hadron structure function at small \textit{x} may proceed faster if the highly packed regions (hot spots) of proton have fractal structures. However, these ideas received wider attention in 2002 when Lastovicka \cite {7, 8} proposed a relevant formalism and a functional form of the structure function $ F_{2}(x,Q^2) $ at small \textit{x}. In recent years, the formalism was further analyzed phenomenologically \cite {9, 10, 11}.\\

The study of structure functions at small \textit{x} has become topical in view of the high energy colliders like HERA \cite {12} and LHC \cite {13, 14} where previously unexplored small \textit{x} regime is being reached. In fixed-target DIS experiments scaling violations have been observed, i.e. the variation at fixed values of Bjorken-\textit{x} of the structure functions with $ Q^{2} $, the squared four-momentum transfer between lepton and nucleon. These scaling violations of $ F_{2}(x,Q^{2}) $ are well described by the DGLAP evolution equations \cite {15, 16, 17}. The strong scaling violations observed at low \textit{x} are attributed to the high gluon density in the proton. In the Quark Parton Model (QPM), $ F_{2}(x,Q^{2}) $ is the sum of the quark and anti-quark \textit{x} distributions, weighted by the square of the electric quark charges, i.e.
\begin{equation}
F_{2}(x,Q^2)= x\sum_{i}e_{i}^2(q_{i}(x,Q^2)+\bar{q_{i}}(x,Q^2))
\end{equation}

The formalism of Ref. \cite{7, 8} was based on unintegrated parton density $ f_{i}(x,k_{t}^{2}) $, or presently more familiar, Transverse Momentum Dependent Parton Density Function TMDPDF \cite {18}, where $ k_{t}^{2} $ is the parton transverse momentum squared. Structure function $ F_{2}(x,Q^{2}) $ was then obtained by using the relation
\begin{equation}
q_{i}(x,Q^2)=\int\limits_0^{Q^{2}} \, dk_{t}^{2} \; f_{i}(x,k_{t}^{2})
\end{equation}

As TMDPDF $ f_{i}(x,k_{t}^{2}) $ is by construction dimensionless, the integrated one $ q_{i}(x,Q^{2}) $ becomes dimension of inverse area contrary to the common expectation.\\

One of the main objectives of the present paper is to study if this formalism can be reformulated by using the standard relation \cite {19, 20, 21}:
\begin{equation}
q_{i}(x,Q^2)=\int\limits_0^{Q^{2}} \, \frac {dk_{t}^{2}}{k_{t}^{2}} \; f_{i}(x,k_{t}^{2})
\end{equation}
instead of Eq (2) so that the integrated quark density also becomes dimensionless.\\

One of the boundary conditions of the structure function $ F_{2}(x,Q^{2}) $ is that it should vanish as $ x\rightarrow1 $ \cite {22, 23, 24}, i.e.
\begin{equation}
\lim_{x \rightarrow 1}F_{2}(x,Q^2)=0
\end{equation}

In the present work we will explore if the Fractal Inspired Model of Ref. \cite{7, 8} is compatible with this condition. We will show that this boundary condition is not physically possible because the function becomes indeterminate. We will therefore go back to the original TMDPDF and modify it by introducing a new function $ h(x,k_{t}^{2}) $ in it.\\

We also explore the possibility of incorporation of saturation of Froissart-Martin limit \cite {25, 26} within this approach as it has attracted attention in the recent literature \cite {27, 28, 29}. It is well-known that in the conventional QCD evolution equations like DGLAP \cite {15, 16, 17} and BFKL \cite {30, 31, 32}, this limit is violated; while in DGLAP approach the small \textit{x} gluons grow faster than any power of $ \ln 1/x \approx \ln (s/Q^{2}) $ \cite {22}, in BFKL it grows as power of $ 1/x $ \cite {30, 31, 32, 33}.\\

The study of the Froissart-Martin bound in DIS is more complicated than that of the hadron-hadron case because we have the external virtual photon mass $ Q $ as an additional kinematic variable. A remarkable modification of the bound in DIS analysis has been suggested by Gribov \cite {34}. Later Gotsman, Levin and Maor \cite {35} proposed a generalization of the Gribov formula. We will comment on the implication of such generalization in the TMDPDFs.\\

In the Ref. \cite {7, 8} the basic TMDPDFs have only one hard scale $ k_{t}^{2} $ (as in BFKL approach) but in general, it may have two hard scales: $ k_{t}^{2} $ and the photon virtuality $ Q^{2} $ \cite {20} as in the CCFM approach \cite {36, 37}.\\

In this paper, we will also generalize the formalism of Ref. [7, 8] with TMDPDFs having two hard scales.
\section{Formalism}
\subsection{Self-similarity based TMDPDFs with One Hard Scale}

One of the most intriguing problems in QCD is the growth of the cross-sections for hadronic interactions with energy. Conventionally deep inelastic lepton-hadron scattering is described in terms of scale-dependent parton distributions $ xg(x,k_{t}^{2}) $ and $ xq(x,k_{t}^{2}) $. These distributions correspond to the density of partons in the proton with longitudinal momentum fraction \textit{x} and parton transverse momentum $ k_{t}^{2} $ \cite {17}.\\

The self-similarity based Model of the nucleon structure function proposed in Ref. \cite {7, 8} has been designed to be valid at small Bjorken-\textit{x}. The formalism described in these references is based on the imposition of self-similarity constraints to the dimensionless TMD quark density $f_{i}(x,k_{t}^{2}) $ and relate it to the integrated density. In other words, using magnification factors $ \displaystyle \frac{1}{x} $ and $ \displaystyle \left( 1+\frac{k_{t}^{2}}{Q_{0}^{2}}\right) $, an unintegrated quark density (TMD) is given as:\\
\begin{equation}
\log f_{i}(x,k_{t}^{2})=D_{1}\log \frac{1}{x}\log \left(1+\frac{k_{t}^{2}}{Q_{0}^{2}}\right)+D_{2}\log \frac{1}{x}+D_{3}\log \left(1+\frac{k_{t}^{2}}{Q_{0}^{2}}\right)+D_{0}^{i}
\end{equation}
where \textit{i} denotes a quark flavor. Here, $ D_{2} $ and $ D_{3} $ are the fractal parameters; $ D_{1} $ is the dimensional correlation relating the two magnification factors; while $ D_{0}^{i}$ is the normalization constant. Conventional integrated quark densities [PDF] $ q_{i}(x,Q^{2}) $ are defined as sum over all contributions with quark virtualities smaller than that of the photon probe $ Q^{2} $. Thus $ f_{i}(x,k_{t}^{2}) $ has to be integrated over $ k_{t}^{2} $ (Eq (2)) to obtain $ q_{i}(x,Q^{2})$.\\

The following analytical parameterization of a quark density is obtained by using Eq (2).
\begin{equation}
q_{i}(x,Q^{2})=\frac{e^{D_{0}^{i}}Q_{0}^{2}x^{-D_{2}}}{1+D_{3}+D_{1}\log \frac{1}{x}}\left( x^{-D_{1}\log \left(1+\frac{Q^{2}}{Q_{0}^{2}}\right)}\left( 1+\frac{Q^{2}}{Q_{0}^{2}}\right)^{D_{3}+1}-1 \right) 
\end{equation}
Now using Eq (6) in Eq (1), the expression for proton structure function is obtained as:\\
\begin{equation}
F_2(x,Q^{2})=\frac{e^{D_{0}}Q_{0}^{2}x^{-D_{2}+1}}{1+D_{3}+D_{1}\log \frac{1}{x}}\left( x^{-D_{1}\log \left(1+\frac{Q^{2}}{Q_{0}^{2}}\right)}\left( 1+\frac{Q^{2}}{Q_{0}^{2}}\right)^{D_{3}+1}-1 \right)  
\end{equation}
where $ D_{0}=\Sigma_{i} D_{0}^{i} $ .\\

As we have seen above, Lastovicka's Model starts with self-similar TMDPDFs. Imposing large \textit{x} limit, i.e. at $x \rightarrow 1$, Eq (7) becomes:\\
\begin{equation}
F_{2}(1,Q^{2})=\frac{e^{D_{0}}Q_{0}^{2}}{1+D_{3}}\left( \left( 1+\frac{Q^{2}}{Q_{0}^{2}}\right)^{D_{3}+1}-1 \right)
\end{equation}
Let
\begin{equation}
\tilde{F}_{2}(1,Q^{2})=\frac{F_{2}(1,Q^{2})}{e^{D_{0}}Q_{0}^{2}}=\frac{1}{1+D_{3}}\left( \left( 1+\frac{Q^{2}}{Q_{0}^{2}}\right)^{D_{3}+1}-1 \right)
\end{equation}
\\
If $ F_{2}(1,Q^{2})=0 $, then we get $ D_{3}=-1 $ and subsequently $ \tilde{F}_{2}(x,Q^{2}) $ becomes indeterminate (using Eq (9)). It implies that the Model does not have correct $ x\rightarrow1 $ behavior. This calls for generalization of PDFs to add a new term in TMDPDFs. A plausible way to incorporate correct $ x\rightarrow1 $ behavior for TMDPDF is to introduce an additional term $ h(x,k_{t}^{2}) $ which does not have fractal basis.\\

Now, starting with the TMDPDF, we have:\\
\begin{eqnarray}
log f_{i}(x,k_{t}^{2})& = & D_{1}\log \frac{1}{x}\log \left(1+\frac{k_{t}^{2}}{Q_{0}^{2}}\right)+D_{2}\log \frac{1}{x}+D_{3}\log \left(1+\frac{k_{t}^{2}}{Q_{0}^{2}}\right)+ \nonumber \\
& & {} D_{0}^{i}+\log h(x,k_{t}^{2})
\end{eqnarray}
i.e.
\begin{equation}
f_{i}(x,k_{t}^{2})=\left(\frac{1}{x} \right)^{D_{1}\log\left( 1+\frac{k_{t}^{2}}{Q_{0}^{2}} \right)}\left(\frac{1}{x} \right)^{D_{2}}\left( 1+\frac{k_{t}^{2}}{Q_{0}^{2}}\right)^{D_{3}} e^{D_{0}} h(x,k_{t}^{2})
\end{equation}
Thus the integrated quark density is given as:\\
\begin{equation}
q_{i}(x,Q^{2})=\int\limits_0^{Q^{2}}\,dk_{t}^{2}\left(\frac{1}{x} \right)^{D_{1}\log\left( 1+\frac{k_{t}^{2}}{Q_{0}^{2}} \right)}\left(\frac{1}{x} \right)^{D_{2}}\left( 1+\frac{k_{t}^{2}}{Q_{0}^{2}}\right)^{D_{3}} e^{D_{0}^{i}} h(x,k_{t}^{2})
\end{equation}
\\
In the simplest case, consider $ h(x,k_{t}^{2}) $ to be only \textit{x}-dependent, i.e. $ h(x,k_{t}^{2}) \approx h(x) $.\\
Then, the improved model is
\begin{equation}
F_{2}^{I}(x,Q^{2})=F_{2}(x,Q^{2}).h(x);\qquad \qquad where \quad 0\leq x \leq 1
\end{equation}
where $ F_{2}(x,Q^{2}) $ is the original Lastovicka's model (Eq (7)).\\

Neglecting $ k_{t}^{2} $ dependence of $ h(x,k_{t}^{2}) $, the most economical way of having correct large \textit{x} behavior is $ h(x,k_{t}^{2}) \approx h(x) \sim (1-x)^{\alpha} , \mathrm{for} \; \alpha > 0 $, but in general the integration is to be carried out with its explicit $ k_{t}^{2} $ dependence in Eq (12). Thus large \textit{x} effect should be introduced as an additional component of TMDPDF to make the PDF compatible with the theoretical expectation, i.e. $ F_{2}(x,Q^{2})\rightarrow0 $ as $ x\rightarrow1 $ \cite {22, 23, 24}.
\subsection{Self-similarity based TMDPDFs with Two Hard Scales}
In general, TMDPDF depends on two hard scales, parton transverse momentum $ k_{t}^{2}$ and the scale $ Q^{2} $ of the probe \cite {20, 38, 39}. In Fractal Inspired Model (Lastovicka's), it will therefore have three magnification factors, viz. $ \displaystyle \frac{1}{x} $, $ \displaystyle \left(1+\frac{k_{t}^{2}}{k_{0}^{2}}\right)$ and $ \displaystyle \left(1+\frac{Q^{2}}{Q_{0}^{2}}\right)$; and four correlation terms. The generalization of TMDs of this Model is therefore given as:
\begin{eqnarray}
\lefteqn{log f_{i}(x,k_{t}^{2},Q^{2}) = } \nonumber \\
& & D_{1}\log \frac{1}{x}\log \left(1+\frac{k_{t}^{2}}{k_{0}^{2}}\right)+D_{2}\log\frac{1}{x}+D_{3}\log \left(1+\frac{k_{t}^{2}}{k_{0}^{2}}\right)+ \nonumber \\
& & D_{4}\log \left(1+\frac{Q^{2}}{Q_{0}^{2}}\right)+ D_{5}\log \frac{1}{x}\log \left(1+\frac{Q^{2}}{Q_{0}^{2}}\right)+ \nonumber \\
& & D_{6}\left(1+\frac{k_{t}^{2}}{k_{0}^{2}}\right)\left(1+\frac{Q^{2}}{Q_{0}^{2}}\right)+ D_{7}\log \frac{1}{x}\log \left(1+\frac{k_{t}^{2}}{k_{0}^{2}}\right)\left(1+\frac{Q^{2}}{Q_{0}^{2}}\right)+ \nonumber \\
& & D_{8}\log h\left(x,k_{t}^{2},Q^{2}\right)+D_{0}^{i}
\end{eqnarray}
\\
where $ h(x,k_{t}^{2},Q^{2}) $accounts for large \textit{x} behavior.\\

In the above Eq (14), there are altogether eight parameters, viz.$ D_{2} $, $ D_{3} $, $ D_{4} $ (\textit{Fractal dimensions});$ D_{1} $, $ D_{5} $, $ D_{6} $, $ D_{7} $ (\textit{Correlation parameters}) and $D_{0}^{i} $ (\textit{Normalization constant}); as well as one unknown function $ h(x,k_{t}^{2},Q^{2}) $.\\

While $ D_{1} $, $ D_{5} $, $ D_{6} $, $ D_{7} $ are the correlation parameters correlating the different magnification factors, $ h(x,k_{t}^{2},Q^{2}) $ is the additional function that cannot be constrained by the notion of fractal geometry.\\
Now following the same steps as above, Eq (14) can be rewritten in the following way:
\begin{eqnarray}
f_{i}(x,k_{t}^{2},Q^{2})& = & e^{D_{0}^{i}} . h(x,k_{t}^{2},Q^{2})\left(\frac{1}{x}\right)^{D_{2}} \left(1+\frac{k_{t}^{2}}{k_{0}^{2}}\right)^{D_{3}} \left(1+\frac{Q^{2}}{Q_{0}^{2}}\right)^{D_{4}}. \nonumber \\
& & {} \left(\frac{1}{x}\right)^{D_{1}\log \left( 1+\frac{k_{t}^{2}}{k_{0}^{2}}\right)} \left(\frac{1}{x}\right)^{D_{5}\log \left( 1+\frac{Q^{2}}{Q_{0}^{2}}\right)} \left(1+\frac{k_{t}^{2}}{k_{0}^{2}}\right)^{D_{6}\log \left( 1+\frac {Q^{2}}{Q_{0}^{2}}\right)}. \nonumber \\
& & {} \left( \frac{1}{x}\right)^{D_{7}\log \left(1+\frac {k_{t}^{2}}{k_{0}^{2}}\right)\log \left(1+\frac {Q^{2}}{Q_{0}^{2}}\right)}
\end{eqnarray}
The integrated quark density is then given as:\\
\begin{equation}
q_{i}(x,Q^{2})=e^{D_{0}^{i}} \left( \frac{1}{x}\right)^{D_{2}+D_{5}\log \left( 1+\frac{Q^{2}}{Q_{0}{2}}\right)}\left( 1+\frac{Q^{2}}{Q_{0}^{2}}\right)^{D_{4}} I(x,Q^{2})
\end{equation}
where\\
\begin{eqnarray}
I(x,Q^{2})& = & \int\limits_0^{Q^{2}}\,dk_{t}^{2}\; h(x,k_{t}^{2},Q^{2}) \left(\frac{1}{x} \right)^{D_{1}\log \left(1+\frac{k_{t}^{2}}{k_{0}^{2}}\right)+D_{7}\log \left( 1+\frac{k_{t}^{2}}{k_{0}^{2}}\right)\log \left(1+\frac{Q^{2}}{Q_{0}^{2}} \right)} . \nonumber \\
& & {} \left( 1+\frac{k_{t}^{2}}{k_{0}^{2}}\right)^{D_{3}+D_{6}\log \left(1+\frac{Q^{2}}{Q_{0}^{2}}\right)}
\end{eqnarray}
Eq (16) is the most general form of PDF which contains two cut-off parameters: $ k_{0}^{2} $ and $ Q_{0}^{2} $ .\\

In a special case, imposing the limit $ D_{4} $, $ D_{5} $, $ D_{6} $, $ D_{7}=0 $ and $ D_{8}=0 $ (i.e. $ h(x,k_{t}^{2},Q^{2})=1 $ [see Eq (14)]), the above integral of Eq (17) becomes:\\
\begin{equation}
I(x,Q^{2})=\int\limits_0^{Q^{2}}\,dk_{t}^{2}\left( \frac{1}{x}\right)^{D_{1}\log \left( 1+\frac{k_{t}^{2}}{k_{0}^{2}}\right)}\left( 1+\frac{k_{t}^{2}}{k_{0}^{2}}\right)^{D_{3}}
\end{equation}
which is the original Lastovicka's Integral to obtain Eq (6).
Now let us proceed to evaluate the integral of Eq (17), in more general case, with the following observation:\\
The exponent of $ \displaystyle \frac {1}{x} $ in Eq (17) can be re-expressed as \\
$ \log \left(1+\frac{k_{t}^{2}}{k_{0}^{2}}\right) \tilde{D}_{1} $ with $ \tilde{D}_{1}= \left\lbrace D_{1}+D_{7}\log \left(1+\frac{Q^{2}}{Q_{0}^{2}}\right) \right\rbrace $ .\\

Similarly, the exponent of $ \displaystyle \left( 1+\frac{k_{t}^{2}}{k_{0}^{2}}\right) $ in Eq (17) is $ \tilde{D}_{3}=D_{3}+D_{6}\log \left( 1+\frac{Q^{2}}{Q_{0}^{2}}\right) $.\\

Without explicit form of $ h(x,k_{t}^{2},Q^{2}) $, Eq (17) cannot be evaluated analytically.\\

In the limit $ h(x,k_{t}^{2},Q^{2})=1 $, Eq (17) then becomes:\\
\begin{equation}
I(x,Q^{2})=\int\limits_0^{Q^{2}}\,dk_{t}^{2}\; \left( \frac{1}{x}\right)^{\tilde{D}_{1}\log \left( 1+\frac{k_{t}^{2}}{k_{0}^{2}}\right)}\left( 1+\frac{k_{t}^{2}}{k_{0}^{2}}\right)^{\tilde{D}_{3}}
\end{equation}
\\
which is again similar to Eq (18), i.e. Lastovicka's Integral, except that in this case $ D_{1} $ is replaced by $ \tilde{D}_{1}$ and $ D_{3} $ by $ \tilde{D}_{3}$ . Eq (19) implies\\
\begin{equation}
I(x,Q^{2})=\frac{Q_{0}^{2}}{1+\tilde{D}_{3}+\tilde{D}_{1}\log \frac{1}{x}}\left\lbrace x^{-\tilde{D}_{1}\log \left(1+\frac{Q^{2}}{k_{0}^{2}}\right)}\left( 1+\frac{Q^{2}}{k_{0}^{2}}\right)^{\tilde{D}_{3}+1}-1 \right\rbrace 
\end{equation}
\\\\
Now substituting this integral from Eq (20) in Eq (16), we obtain:\\
\begin{eqnarray}
q_{i}(x,Q^{2})& = & \frac{e^{D_{0}^{i}}\; Q_{0}^{2}\; \left( \frac{1}{x}\right)^{D_{2}+D_{5}\log \left(+\frac{Q^{2}}{k_{0}^{2}} \right)}}{1+\tilde{D}_{3}+\tilde{D}_{1}\log \frac{1}{x}} \left(1+\frac{Q^{2}}{k_{0}^{2}}\right)^{D_{4}}. \nonumber \\
& & {} \left\lbrace x^{-\tilde{D}_{1}\log \left( 1+\frac{Q^{2}}{k_{0}^{2}}\right)}\left(1+\frac{Q^{2}}{k_{0}^{2}}\right)^{\tilde{D}_{3}+1}-1 \right\rbrace 
\end{eqnarray}
The expression for the proton structure function $F_{2}^{I}(x,Q^{2}) $ can thus be obtained using Eq (21) in Eq (1):
\begin{eqnarray}
F_{2}^{I}(x,Q^{2})& = & \frac{e^{D_{0}}\; Q_{0}^{2}\; \left( \frac{1}{x}\right)^{D_{2}+D_{5}\log \left(+\frac{Q^{2}}{k_{0}^{2}} \right)-1}}{1+\tilde{D}_{3}+\tilde{D}_{1}\log \frac{1}{x}}\left(1+\frac{Q^{2}}{k_{0}^{2}}\right)^{D_{4}}. \nonumber \\
& & \left\lbrace x^{-\tilde{D}_{1}\log \left( 1+\frac{Q^{2}}{k_{0}^{2}}\right)}\left(1+\frac{Q^{2}}{k_{0}^{2}}\right)^{\tilde{D}_{3}+1}-1 \right\rbrace
\end{eqnarray}
where $ \tilde{D}_{1} $ and $ \tilde{D}_{3} $ are independent of $k_{t}^{2}$ .\\

As in Eq (7), imposing boundary condition (Eq (4)) at $ x\rightarrow1 $, Eq (22) yields $\tilde{D}_{3}=-1$ and subsequently $ F_{2}^{I}(x,Q^{2}) $ becomes indeterminate.\\

Thus the correct $ x\rightarrow1 $ limit of the structure function (Eq (22)) cannot be incorporated as in the original one. Therefore the most economic way to incorporate large \textit{x} behavior is by putting $ h(x,k_{t}^{2})\neq 1 $ but of the factorizable form $ h(k_{t}^{2})(1-x)^{\alpha}$ so that it vanishes as $ x\rightarrow1 $. As in the original Model it can be seen that $ x\rightarrow1 $ limit of Eq (22) with $ h(x,k_{t}^{2})\sim h(k_{t}^{2})(1-x)^{\alpha}, \mathrm{for} \; \alpha > 0 $, exists. As noted earlier (with Eq (13)), we have neglected explicit $k_{t}^{2}$ (and $Q^{2}$ as well) dependence of $ h(x,k_{t}^{2},Q^{2}) $ in the most economical version of the model. However, as noted above, the integration of Eq (17) can be carried out only with its explicit $ k_{t}^{2} $ dependence.
\subsection{Problem of Dimensionality}
Exact relation between TMDPDF and PDF is still an active area of study \cite {38, 39, 40}. Formally, the integration over the transverse component of the parton momentum is expected to yield the PDF (Eq (2)). However, the self-silmilar TMDPDF (Eq (5), Eq (10) and Eq (14)) are by construction dimensionless in the formalism. As a result, the resulting PDF as well as the proton structure function possess dimension of inverse area. Literally speaking, PDF has become dimensional while TMDPDF is dimensionless. Phenomenologically, it matters not much but in principle, it is questionable. If instead we consider Eq (3), $ k_{t}^{2}$ in the denominator plays the role of dimensionality check. But even if dimensionality problem is solved, there arises a singularity at $ k_{t}^{2}=0 $. This can be overcome by introducing a minimum transverse cut-off momentum. The crucial point is that in order to make the PDF finite and dimensionless in the formalism, we need to have a cut-off over $ k_{t}^{2}$, say $ \tilde{k}_{0}^{2}$.\\
Thus, the dimensionless Lastovicka's integral can be written as:
\begin{equation}
\hat{I}(x,Q^{2})=\int\limits_{\tilde{k}_{0}^{2}}^{Q^{2}}\,\frac{dk_{t}^{2}}{k_{t}^{2}}\left( \frac{1}{x}\right)^{D_{1}\log \left( 1+\frac{k_{t}^{2}}{k_{0}^{2}}\right)}\left( 1+\frac{k_{t}^{2}}{k_{0}^{2}}\right)^{D_{3}}
\end{equation}
So the dimensionless PDF in the original Lastovicka's Model becomes:\\
\begin{equation}
q_{i}^{*}(x,Q^{2})=e^{D_{0}^{i}}\left(\frac{1}{x} \right) ^{D_{2}} \hat{I}(x,Q^{2})
\end{equation}
and the corresponding structure function becomes:\\
\begin{equation}
F_{2}^{*}(x,Q^{2})=e^{D_{0}}\left(\frac{1}{x} \right)^{D_{2}-1} \hat{I}(x,Q^{2})
\end{equation}
\\
to be compared with Eq (6) and Eq (7) respectively.\\

The formalism can be generalized to TMDPDF with two hard scales as well. In this case, dimensionless PDF in the two-scale generalized Lastovicka's Model is given as:\\
\begin{equation}
\tilde{q}_{i}(x,Q^{2})=e^{D_{0}^{i}}\left(\frac{1}{x}\right)^{D_{2}+D_{5}\log \left( 1+\frac{Q^{2}}{Q_{0}^{2}}\right)}\left(1+\frac{Q^{2}}{Q_{0}^{2}}\right)^{D_{4}}\tilde{I}(x,Q^{2})
\end{equation}
where
\begin{equation}
\tilde{I}(x,Q^{2})=\int\limits_{\tilde{k}_{0}^{2}}^{Q^{2}}\,\frac{dk_{t}^{2}}{k_{t}^{2}}\left( \frac{1}{x}\right)^{\tilde{D}_{1}\log \left( 1+\frac{k_{t}^{2}}{k_{0}^{2}}\right)}\left( 1+\frac{k_{t}^{2}}{k_{0}^{2}}\right)^{\tilde{D}_{3}}
\end{equation}
\\
and consequently, the structure function has the form:\\
\begin{equation}
\tilde{F}_{2}(x,Q^{2})=e^{D_{0}}\left( \frac{1}{x}\right)^{D_{2}+D_{5}\log \left(1+\frac{Q^{2}}{Q_{0}^{2}}\right)-1}\left(1+\frac{Q^{2}}{Q_{0}^{2}}\right)^{D_{4}}\tilde{I}(x,Q^{2})
\end{equation}
\\
to be compared with Eq (21) and Eq (22) respectively.
\subsection{Froissart Bound and Self-Similarity}
Froissart bound \cite {25, 26} implies that the total cross-section (or equivalently structure functions) cannot rise faster than the logarithmic growth of $ \ln^{2}\displaystyle \left(\frac{1}{x}\right) $. Such a slow logarithmic growth is however not compatible with the notion of fractal dimension as occurred in the original Fractal Inspired Model \cite{7, 8}. The reason is that self-similar models allow only power law growth and not the logarithmic one. This statement can be explained in a more explicit way in the following manner.\\

\textbf{(i) Power law growth}\\

Let $ N(R) $ be the number of self-similar objects, $ R(N) $ be the magnification factor and $ D $ the exponent of the \textit{power law} behavior, then the relation is given as:
\begin{equation}
N(R)=cR(N)^{-D}
\end{equation}
where \textit{c} is the proportionality constant. Eq (29) leads to\\
\begin{equation}
D\approx \lim_{N \to \infty , R \to 0} \frac{\log N(R)}{\log \left( 1/R(N)\right)}
\end{equation}
\\
Eq (30) shows that \textit{D} is ultimately independent of $ N(R) $ and $ R(N) $ in the limit $ N\rightarrow \infty $ and $ R\rightarrow0 $. It implies that \textit{D} is universal property of a self-similar object to be identified as its fractal dimension \cite {41}.\\

\textbf{(ii) Logarithmic growth}\\

If, on the other hand, number of self-similar objects is assumed to grow logarithmically, i.e.\\
\begin{equation}
N(R)=\frac{c}{(R(N))^{D}}
\end{equation}
one obtains
\begin{equation}
D_{N \to \infty}\approx -\frac{\ln N(R)}{\ln \ln R(N)}
\end{equation}
\\
Eq (32) implies that limiting \textit{D} is not $ N(R) $ or $ R(N) $-independent and hence it is not an intrinsic property of the object. So it cannot be identified as fractal dimension.\\
Thus the logarithmic growth and hence the Froissart behavior cannot be described in a self-similar model.

\subsection{Froissart Bound and TMDPDFs}
Understanding the behavior of high energy hadron reactions from a fundamental perspective within QCD is an important goal of high energy physics. It was already observed that the total hadronic cross-section grows with the centre of mass energy $ (\sqrt{s}) $. A QCD based explanation for this rising behavior was proposed by Gaisser and Halzen \cite {42}: the cross-section would grow because partons would begin playing a dominant role in the hadronic reactions. At high energy, the parton distribution, especially the gluon distribution, grows very rapidly and thus leads to rising cross-section. It is expected that at very high energies the hadronic cross-sections satisfy the Froissart bound, which is a well-established property of the strong interactions \cite {29}. It has been shown that the Froissart bound is saturated at very high energies \cite {43, 44} in $ \gamma p $, $ \pi^{\pm} p $ and $\bar{p} p$ \& $ pp $ scattering. Saturation of the Froissart bound refers to an energy dependence of the total cross-section rising no more rapidly than $ \ln^{2}s $ \cite {27}.\\

It is already known that the BFKL evolution equation \cite {30, 31, 32} sums leading logarithms $ \ln (1/x) $. The gluon distribution function, governed by this equation, is found to increase at small \textit{x} and it has been confirmed by the recent HERA data of the proton structure function $ F_{2}(x,Q^{2}) $ which is involved in DIS \cite {45, 46}. The increase is however power-like, such that $ F_{2}(x,Q^{2}) $ and the DIS cross-section rise as power of $ \left( \frac{1}{x}\right) $, i.e., as power of \textit{s} for $ x=Q^{2}/s $. This behavior does not satisfy the Froissart bound and thus violates unitarity \cite {47}. The BFKL equation cannot thus be the final theory for small \textit{x} physics.\\

This bound \cite {25, 26} derived from analyticity and unitarity demands that $ F_{2}(x,Q^{2}) $ grow no more rapidly than $ \displaystyle \ln^{2} \left( \frac{1}{x}\right)  $ at very small \textit{x} for all values of $ Q^{2} $. Therefore, for fixed $ Q^{2} $ and small \textit{x}, if the condition\\
\begin{equation}
F_{2}(x,Q^{2})h(x,k_{t}^{2})\leq \ln ^{2}\left( \frac{1}{x}\right)
\end{equation}
\\
is satisfied, then the Model satisfies the Froissart bound with $ F_{2}(x,Q^{2}) $ and $ h(x,k_{t}^{2}) $ as defined in Eq (7) and Eq (11).\\

Consider $ F_{2}(x,Q^{2}) $ at $ Q^{2}=Q_{0}^{2} $ and ultra small \textit{x}. Then, $ D_{1}\log \left(\frac{1}{x} \right)>> D_{3},1 $ and therefore Eq (7) yields\\
\begin{equation}
F_{2}(x,Q_{0}^{2})\approx \frac{\left( \frac{1}{x}\right)^{D_{2}+D_{1}\ln 2 -1}}{\ln \left( \frac{1}{x}\right)}
\end{equation}
Using Eq (34) in Eq (33), we have:\\
\begin{equation}
h(x,Q_{0}^{2}) \lesssim \frac{\ln ^{3}\left(\frac{1}{x} \right) }{\left( \frac{1}{x}\right)^{D_{2}+D_{1}\ln 2 -1} }
\end{equation}
for a given $ k_{t}^{2} $.\\

Eq (35) gives the specific limiting behavior of $ h(x,Q^{2}) $ for Froissart bound compatibility of Eq (7) at $ Q^{2}=Q_{0}^{2} $.\\

Froissart compatibility condition (Eq (35)) can be generalized to any $ Q^{2} $ for ultra small \textit{x}.
\begin{equation}
h(x,Q^{2})\lesssim \frac{\ln ^{3}\left(\frac{1}{x}\right)}{\left( \frac{1}{x}\right)^{D_{2}-1}\left( \frac{1}{x}\right)^{D_{1}\log \left( 1+\frac {Q^{2}}{Q_{0}^{2}}\right)}\left(1+\frac {Q^{2}}{Q_{0}^{2}} \right)^{D_{3}+1}}
\end{equation}
\\
In Ref. [35], Gotsman, Levin and Maor had improved Froissart-Martin bound for ultra small \textit{x} at fixed $ Q^{2}$ for virtual photon-proton cross-section:\\
\begin{equation}
\sigma(r^{\ast},p)\approx \frac{F_{2}^{p}(x,Q^{2})}{1-x} \lesssim \left( \ln ^{5/2}\frac{1}{x}\right) 
\end{equation}
\\
In that case the corresponding improved Froissart compatibility condition on the TMDPDF factor $ h(x,k_{t}^{2})$ will be\\
\begin{equation}
h(x,k_{t}^{2})\lesssim \frac{\ln ^{3}\left(\frac{1}{x}\right)}{\left( \frac{1}{x}\right)^{D_{2}+D_{1}\ln 2-1}}
\end{equation}
for any given $ k_{t}^{2}$ at $ Q^{2} = Q_{0}^{2} $.
\subsection{Graphical Representation of Self-similar TMDPDFs}
Specific models of TMDPDFs available in recent literature are summarized in Ref. \cite {48}. Here we analyze the ones of Ref. \cite{7, 8} alone. The TMDPDFs $ f_{i}(x,k_{t}^{2})$ of the original Lastovicka's model (Eq (5)) contain the parameters $ D_{0}^{i} $, $ D_{1} $, $ D_{2} $, $ D_{3} $ and $ Q_{0}^{2} $. These are determined from HERA data in the range:
\begin{eqnarray}
0.045 \leq Q^{2} \leq 150 \; \mathrm{GeV}^{2} \nonumber \\
\mathrm{and} \quad 6.2 \times 10^{-7} \leq x \leq 0.2
\end{eqnarray}
where
\begin{eqnarray}
D_{0} & = & 0.339\pm 0.145 \nonumber \\
D_{1} & = & 0.073\pm 0.001 \nonumber \\
D_{2} & = & 1.013\pm 0.01 \nonumber \\
D_{3} & = & -1.287\pm 0.01 \nonumber \\
Q_{0}^{2} & = & 0.062\pm 0.01 \; \mathrm{GeV}^{2}
\end{eqnarray}
Using Eq (39) and taking $ n_{f}=4 $, $ D_{0}^{i}\approx 0.085 $ (taking $ D_{0}=\Sigma_{i}D_{0}^{i}=n_{f}D_{0}^{i}$), $ f_{i}(x,k_{t}^{2}) $ is given by Eq (5), viz.
\begin{equation}
f_{i}(x,k_{t}^{2})=e^{D_{0}^{i}}\left( \frac{1}{x}\right)^{D_{2}+D_{1}\ln \left(1+\frac {k_{t}^{2}}{Q_{0}^{2}} \right)\left( 1+\frac {k_{t}^{2}}{Q_{0}{2}}\right)^{D_{3}}}
\end{equation}
In Figure 1, $ f_{i}(x,k_{t}^{2}) $ vs $ k_{t}^{2} $ is shown for representative values of $ \mathrm{(i)}\; x = 10^{-4} \mathrm{(ii)}\; x = 10^{-3} \mathrm{(iii)}\; x = 10^{-2} $ and $ \mathrm{(iv)}\; x = 0.1 $.\\
\begin{figure}[h]
\centering
\includegraphics[width=5in,height=5in]{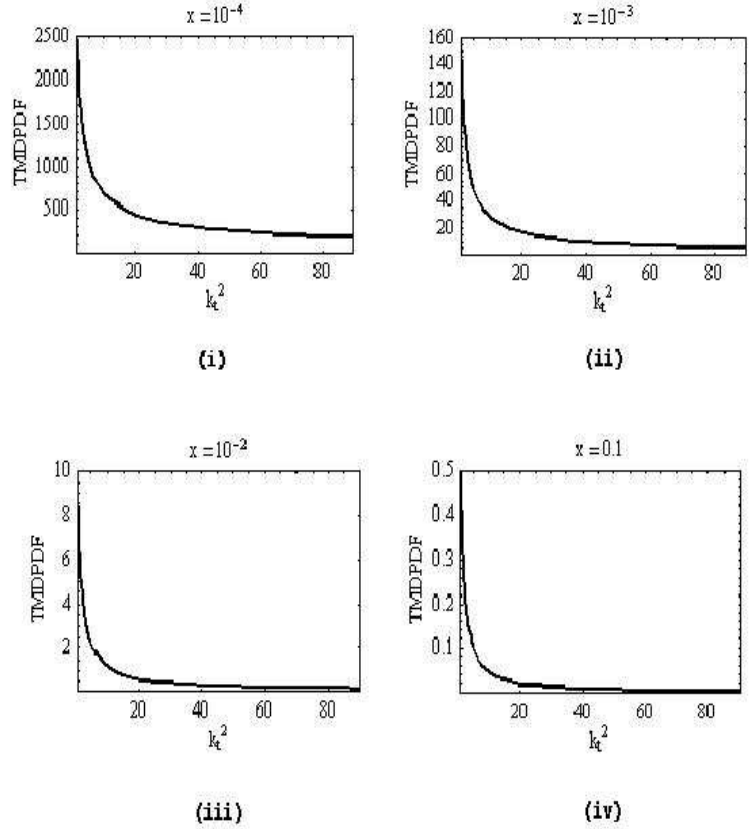}
\caption{$ f_{i}(x,k_{t}^{2}) \; \mathrm{vs} \; k_{t}^{2} $}
\end{figure}
\\\\\\
Similarly, $ f_{i}(x,k_{t}^{2}) $ vs \textit{x} is shown in Figure 2 for representative values of $ \mathrm{(i)}\; k_{t}^{2}=Q_{0}^{2}=0.062 \; \mathrm{GeV}^{2} \mathrm{(ii)}\; k_{t}^{2}=0.5 \; \mathrm{GeV}^{2}  \mathrm{(iii)}\;k_{t}^{2}=1 \; \mathrm{GeV}^{2} $ and $ \mathrm{(iv)}\; k_{t}^{2}=10 \; \mathrm{GeV}^{2} $.\\

\begin{figure}[h]
\centering
\includegraphics[width=5in,height=5in]{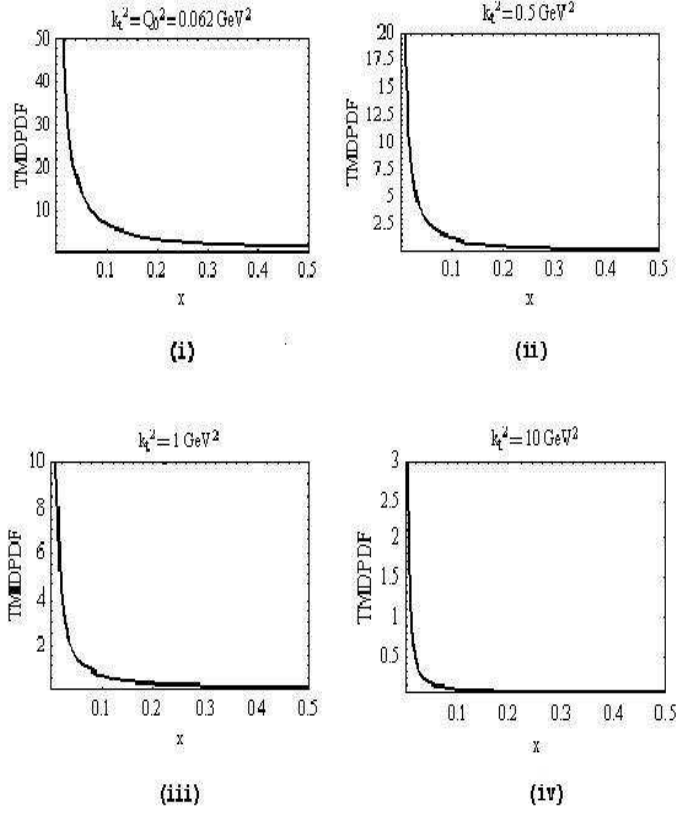}
\caption{$ f_{i}(x,k_{t}^{2}) \; \mathrm{vs} \; x $}
\end{figure}

\section{Conclusion}
In this paper we have generalized the method of parameterization of the self-similar TMDs of Ref. \cite{7, 8} to two hard scales. We have shown that the redefinition of the TMDs is necessary to make them compatible with the theoretical expectation in large \textit{x} limit. We have then studied the compatibility of saturation of the Froissart-Martin bound within this approach and obtained necessary constraints on TMDPDFs for it. Relation between the TMDPDFs and PDFs within such Fractal Inspired Model is also re-examined in the context of their relative dimensions.\\

Let us now compare the present approach with other models based on self-similarity and fractals. In the statistical quark model of Ref. \cite{5}, the proton possesses a fractal structure of fractal dimension 9/2, topological dimension 3 and anomalous dimension 3/2. The present approach falls short of such specific predictions as it has inbuilt multifractal characters. As noted earlier, in the original version of the Fractal Inspired Model \cite{7, 8}, one has two fractal dimensions $ D_{2} $, $ D_{3} $ and one correlation parameter $ D_{1} $ fitted from HERA data. In Ref. \cite {49}, the monofractal limit of the model \cite{7, 8} was indeed studied. Its analysis however indicated that only in a limited $ x-Q^{2} $ range $ (Q^{2} \leq 12 \; \mathrm{GeV}^{2}) $, such notion of monofractality makes sense. In that range, dimensional correlation $ (D_{1}) $ vanishes and the proton possesses fractality $ (D \approx 1.107) $ close to Koch curve $ (D \approx 1.26) $.\\

In Ref. \cite{6}, the proton is viewed as a collection of smaller colorless sources, filling its interior. The density of such sources has fractal structures which gives rise to faster saturation of the proton structure function at small \textit{x}.\\

It is also instructive to recall the physical interpretation of fractal dimension of proton since the notion is rather recent in its description. As is well-known \cite {50}, the fractal dimension measures the way in which distribution of points fill a geometric space on the average. Extending the notion to TMDPDF, fractal dimension tells how densely small \textit{x} partons fill the proton in self-similar way on the average. It is also interesting to note that in the model of Ref. \cite{7, 8}, TMDPDF takes the simplest form\\
\begin{equation}
f(x,k_{t}^{2})\approx \left( \frac{1}{x}\right)^{D}
\end{equation}
in the limit $ D \equiv D_{2} >> D_{1}, D_{3} $.In such a case, fractal dimension is essentially close to \textit{x}-slope \cite {51} or Pomeron intercept \cite {52, 53, 54}.\\

To conclude the problem of self-similarity in the gluon cascades is an interesting one and worth investigating. In this paper, we have used fractality only as a tool to provide parameterization for the TMDPDFs having one and two hard scales. These result in a set of parameters to be fitted from data, as was the case with the original model of Ref. \cite{7, 8}. However, in spite of this inherent limitation, it has been shown how large \textit{x} effects and high density QCD phenomenon like Froissart saturation should appear in such fractal inspired TMDPDFs. Naturally the most interesting study would be to calculate the fractal parameters from the first principles, say from a self-similar Lagrangian \cite {55}. Interesting though, the solution appears to be a formidable task, beyond the reach of the present authors.
\section*{Acknowledgement}
One of the authors (DKC) gratefully acknowledges numerous correspondences with Dr T Lastovicka while the other (AJ) thanks the UGC-RFSMS for financial support.

\end{document}